\documentclass{svproc}
\usepackage{url}

\usepackage{graphics}	
\usepackage{graphicx}	
\usepackage{subfig}
\usepackage{caption}
\usepackage{multicol}

\begin{document}
\mainmatter              
\title{A Proof-of-­principle for Time-Of-Flight ­Positron Emission Tomography Imaging}
\titlerunning{TOF-PET Imaging}  
%
\author{Rajesh Ganai\inst{1,2} \and Shaifali Mehta\inst{3} \and Mehulkumar Shiroya\inst{4} \and Mitali Mondal\inst{1,2}
Zubayer Ahammed\inst{1,2} \and Subhasis Chattopadhyay\inst{1,2}}
\authorrunning{Rajesh Ganai et al.} 
%

%
\institute{EHEP\&A Group, Variable Energy Cyclotron Centre, Kolkata-700064, India,\\
\email{rajesh.ganai.physics@gmail.com}
\and
Homi Bhabha National Institute, Training School Complex, Anushakti Nagar, Mumbai - 400 085, India.
\and
School of Physics and Material Sciences, Thapar University, Patiala, Punjab-147004, India
\and
Sardar Vallabhbhai National Institute of Technology, Surat-395007, Gujarat, India.}

\maketitle              
\vspace{-5mm}
\begin{abstract}
Time-Of-Flight (TOF) is a noble technique that is used in Positron Emission Tomography (PET) imaging worldwide.  Scintillator ­based imaging system 
that is being used around the world for TOF-­PET is very expensive. Multi-gap Resistive Plate Chambers (MRPCs) are gaseous detectors which are 
easy to fabricate, inexpensive and have excellent position and timing resolution. They can be used as a suitable alternative to highly expensive 
scintillators. For the sole purpose of TOF-­PET,  pair of 18 cm $\times$ 18 cm, 5 ­gap, glass ­based MRPC modules have been fabricated. Our main aim was 
to determine the shift in the position of source (Na-­22) with these fabricated MRPCs. In this document the details of the experimental results will 
be presented.
\end{abstract}
\vspace{-10mm}
\section{Introduction}
Positron Emission Tomography PET)\cite{pet_1}, is a radio-­tracer, nuclear medicine imaging technique. PET is used to observe metabolic processes in 
the body. The basic principle of PET is detecting a pair of back to back 511 keV photons created by the annihilation of a positron with an electron. 
The positron ­emitter, Fludeoxyglucose ($^{18}$F)
(FDG) which is a radio-­tracer, administered in the body annihilates into a pair 
of 511 keV photons, flying in opposite directions. PET
is both a medical and research tool. It is used heavily in clinical oncology (medical imaging of
tumors and the search for metastases), and for clinical diagnosis of certain diffuse brain diseases
such as those causing various types of dementias. PET is also an important research tool to
map normal human brain and heart function, and support drug development.

\begin{figure}
\centering
  \includegraphics[width=0.7\linewidth]{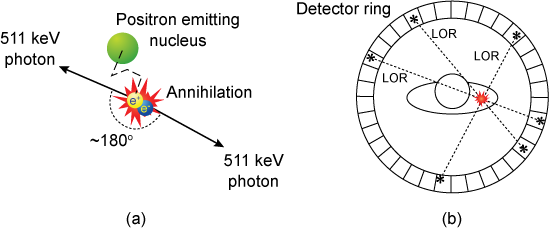}
\captionof{figure}{ \small \sl [Colour online] Imaging principle of PET: (a) After annihilation of a positron and an electron, two 511 keV photons 
are emitted in (almost) opposite directions; (b) When two interactions are simultaneously detected within a ring of detectors surrounding the patient, 
it is assumed that an annihilation occurred on the so-called line-of-response (LOR) connecting the two interactions. By recording many LORs the 
activity distribution can be tomographically reconstructed.}
\label{pet_1}
\end{figure}

Time of flight (TOF) technique has found its application in PET imaging. The two gamma-ray interaction points define a so-called line-of-response 
(LOR) on
which the annihilation must have taken place. A precise measurement of the arrival times of
the coincident photons along with the time difference in flight-time of the two photons helps to localize the annihilation event on the LOR. 
Figure~\ref{pet_1} illustrates the basic working and detection principle of any TOF-PET system.
\vspace{-8mm}
\vspace{-12mm}
\section{Experimental set up and test results}
\begin{figure}[!htb]
\centering
\begin{minipage}{.5\textwidth}
  \centering
  \includegraphics[width=0.8\linewidth]{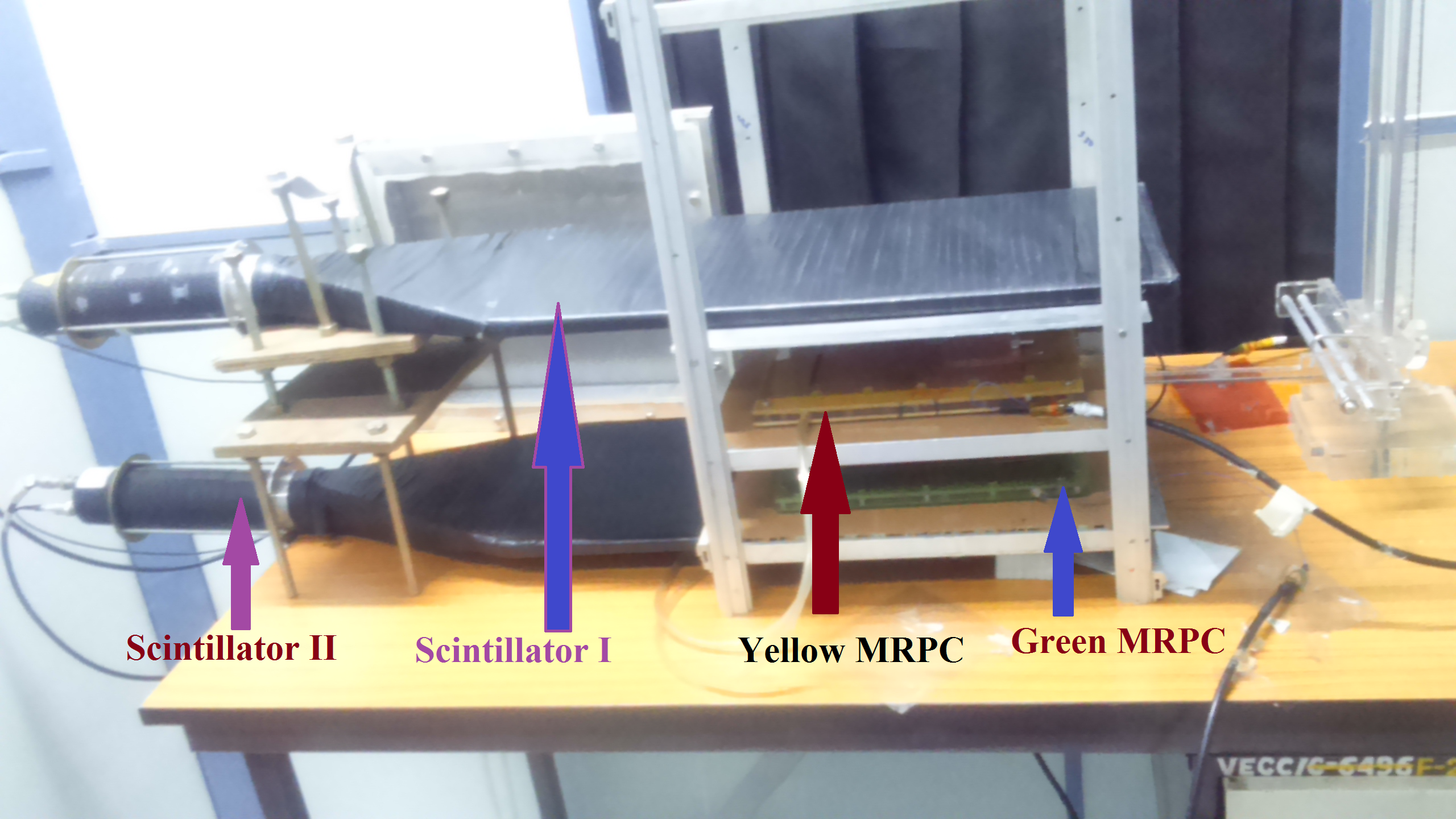}
\end{minipage}%
\begin{minipage}{.5\textwidth}
  \centering
  \includegraphics[width=0.8\linewidth]{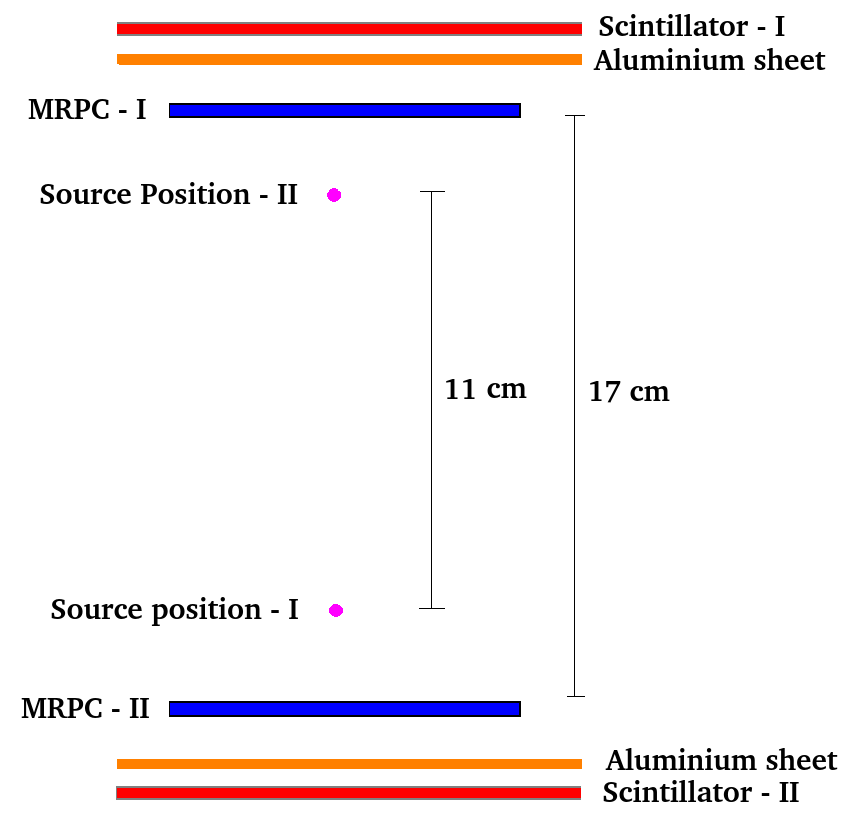}
\end{minipage}
\captionof{figure}{ \small \sl Actual experimental set up and schematic of the experimental set up for TOF-PET with prototype MRPCs.}
\label{pet_set_up}
\end{figure}

\vspace{-2mm}
The aim of this work was to detect the two back to back gammas created by the annihilation of positron emitted from $^{22}$Na with an electron with 
the developed prototype 5-gap glass MRPCs\cite{mrpc_tof_pet_1},\cite{mrpc_tof_pet_2} and also to sense a change in the position of the $^{22}$Na
source. In order to do so, the major challenge was to eliminate the cosmic muon background as MRPCs are known to have very good charged particle 
detection efficiency and was successfully achieved by veto method. The schematic of the experimental set up and the actual set up has been shown in 
figure~\ref{pet_set_up}. The two prototype MRPCs were kept horizontally and separated by a known distance of  17 cm. A $^{22}Na$ source was kept in 
between the MRPCs. Two aluminium plates of dimensions $\sim$ 30 cm $\times$ 30 cm $\times$ 0.5 cm ensured that the back to back photons does not 
reach the scintillators. Two
plastic scintillators each of dimension  $\sim$ 50 cm $\times$ 25 cm were used. 
\vspace{-5mm}

\begin{figure}[!htb]
\centering
\begin{minipage}{.5\textwidth}
  \centering
  \includegraphics[width=0.8\linewidth]{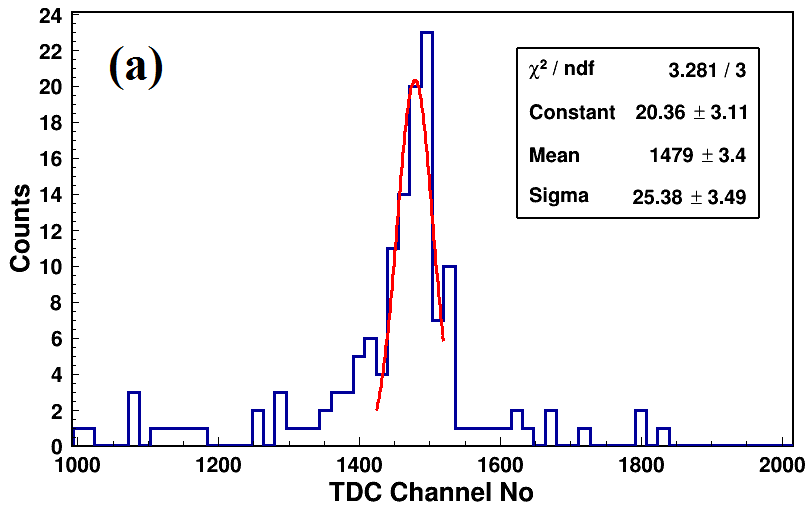}
\end{minipage}%
\begin{minipage}{.5\textwidth}
  \centering
  \includegraphics[width=0.8\linewidth]{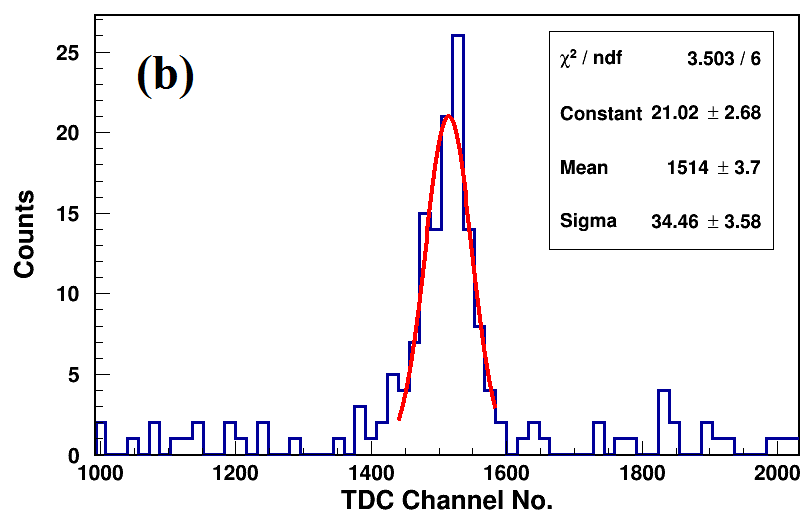}
\end{minipage}
\captionof{figure}{ \small \sl (a) The TDC spectra of MRPC-II when the $^{22}$Na source was kept at position-I. 
(b) The TDC spectra of MRPC-II when the $^{22}$Na source was kept at position-II.
The resolution of TDC is $\sim$25 ps / channel.}
  \label{tdc_spectra}
\end{figure}

A suitable trigger ($ \overline {(Scintillator- I) \cdot (Scintillator - II)}$$\cdot$ MRPC - I) was also chosen which initiated the START of the 
TDC module. The first set of TDC spectra was taken when the source was placed 3 cm away from the
bottom or MRPC - II and the second set was taken when the source was kept at 14 cm away
from the bottom MRPC which have been shown in figure~\ref{tdc_spectra}.
The mean channel of the TDC spectra after a Gaussian fit was obtained to be 1479 for
Figure 3(a) and 1514 for Figure 3(b). Clearly there is a shift in the mean of the TDC spectra,
specially in the mean of the spectra by 35 TDC channels as the source was moved from position-I to position-II by 11 cm. Assuming the velocity of 
photons to be $\sim$30 cm/ns, a change in source position by 11 cm should give TDC channel difference of ∼30 channels which is close to to the obtained 
value of 35 channels. From another way of looking at it, a shift in the mean of the TDC spectra of 35 channels should yield a change in the source 
position by ∼ 12.8 cm which is close to the actual change in source position by 11 cm.
\vspace{-12mm}
\section{Summary}
\vspace{-2mm}
Excellent time resolution of MRPCs make them potential candidate to replace the scintillators in existing PET systems. If successful, the cost per 
scan of PET imaging will reduce drastically as MRPCs are relatively low cost detectors. As a first step towards this noble work, two prototype MRPCs 
have been tested in a two-MRPC coincidence set-up for the detection of back to back photons created by the annihilation of positron (emitted from 
Na$^{22}$ source) with a nearby electron. The change in distance in the source position was successfully estimated from the time spectra obtained by 
using both the MRPCs.
\vspace{-2mm}
%
%

\end{document}